# Quantification of spin accumulation causing spin-orbit torque in Pt/Co/Ta stack

Feilong Luo, Sarjoosing Goolaup, Christian Engel, and Wen Siang Lew[*]

*School of Physical and Mathematical Sciences, Nanyang Technological University,*

*21 Nanyang Link, Singapore 637371*

**Abstract**

Spin accumulation induced by spin-orbit coupling is experimentally quantified in stack with in-plane magnetic anisotropy *via* the contribution of spin accumulation to Hall resistances. Using a biasing direct current the spin accumulation within the structure can be tuned, enabling quantification. Quantification shows the spin accumulation can be more than ten percentage of local magnetization, when the electric current is $10^{11}$ Am$^{-2}$. The spin accumulation is dependent of the thickness of Ta layer, the trend agrees with that of spin Hall angle indicating the capability of Ta and Pt in generating spins.

[*]Corresponding author: wensiang@ntu.edu.sg

**Introduction**

Current-induced spin accumulation causes spin-orbit torque (SOT) on the magnetization of a ferromagnetic metal (FM) layer sandwiched by two heavy metal (HM) layers, *via* exchange interaction [1]. The spin accumulation originates from two spin-orbit coupling effects: Rashba effect and spin Hall effect [2-10]. The SOT is reflected in the revised Landau–Lifshitz–Gilbert equation by the term $-\gamma_0 \mathbf{M} \times J\mathbf{s}$, where $\gamma_0$ is the gyromagnetic coefficient, $\mathbf{M}$ is the magnetization of the FM layer, $\mathbf{s}$ is the spin accumulation, and $J$ is a coefficient related to spin diffusion length of accumulated spins in the FM layer. The term $-\gamma_0 \mathbf{M} \times J\mathbf{s}$ can be decomposed into a fieldlike torque $\boldsymbol{\tau}_F = -H_F \mathbf{M} \times \mathbf{p}$ and a dampinglike torque $\boldsymbol{\tau}_D = -H_D \mathbf{M} \times (\mathbf{m} \times \mathbf{p})$, where $\mathbf{p}$ represents the spin orientation of the electrons diffusing into the FM layer, and $\mathbf{m}$ is the unit vector of $\mathbf{M}$ [1, 10-14]. The corresponding effective fields arising from SOT can be written as the fieldlike term $\mathbf{H}_F = H_F \mathbf{p}$ and dampinglike term $\mathbf{H}_D = H_D \mathbf{m} \times \mathbf{p}$, alternatively, $J\mathbf{s} = H_F \mathbf{p} + H_D \mathbf{m} \times \mathbf{p}$ [6, 8, 10, 12, 15-20]. The effective field, $J\mathbf{s}$, which is a combination of spin accumulation and a spin-diffusion related coefficient, has been widely characterized *via* current-induced domain wall motion [8, 21, 28-30], ferromagnetic resonance (FMR) techniques [31-38], and SOT-assisted magnetization switching [6, 20, 22, 35, 39]. Quantification of the spin accumulation, which plays a crucial role in the origins of the SOT, has remained elusive.

In this letter, we provide a concise solution to quantify the spin accumulation in the sandwiched structure with in-plane magnetic anisotropy (IMA). We propose the spin accumulation *s* contributes to the second harmonic Hall resistance in the harmonic Hall voltage scheme, in addition to the SOT effective field $J$**s** as expected. Applying a biasing direct current (DC) enables the extraction of the contribution of the spin accumulation from the second harmonic Hall resistances. Analogized to first harmonic Hall resistance which is induced by the magnetization, modulation of the second Hall resistance *via* DC current can be used to compute the spin accumulation. Results of the computation show the spin accumulation is dependent of the thickness of HM layers. This quantification allows us to understand the anatomy of $J$**s** and distinguish the roles of $J$ and **s**.

**Main body**

Following the transfer of momentum to the local magnetization, the accumulated spins *s* adopt similar polarization as the magnetization orientation of the FM layer. The structure comprises of Ta/Co/Pt multilayer, where the FM layer exhibits IMA. The initial polarization of *s* is induced by Rashba effect due to the asymmetric HM/FM interface and spin Hall effect within the Ta and Pt layers [2-10, 8, 15, 18, 19, 21-27]. The Rashba effect re-orientates the spin with in the conduction electrons of FM layer to provide a net resultant spin in the FM layer [5]. Additionally, the spin Hall effect induces a spin-selective separation of electrons in the HM layer; the spin polarized electrons then diffuses into the FM layer [10]. In the Co layer, the transfer of spin torque from the spin polarized electron to the FM layer occurs on the nanosecond time scale [5]. A schematic of the spin transfer process is depicted in Fig 1(a). At the end of the spin transfer, *s* is in relaxation state, hence it adopts the same orientation as the local magnetic moment $M$**m** as depicted in Fig. 1(b). In experiment, within the low frequency regime of hundreds of Hertz, corresponding to period of oscillation of alternating current (AC) in millisecond scale, it is reasonable to consider that the accumulated spins $s$ follow the orientation of $M\mathbf{m}$. Similarly, extending to direct current (DC) bias regime, an identical approximation can be made and the spins $s$ similarly aligns along $\mathbf{m}$. Therefore, after the electron spins have transferred the momentum to the local magnetization, the resultant polarization direction of the electron is along the magnetization orientation of the FM layer. Thus, the spin accumulation can be written as $s\mathbf{m}$. Consequently, the total magnetization of the stack becomes to $M\mathbf{m}+s\mathbf{m}$ from $M\mathbf{m}$.

We propose that the spin accumulation $s\mathbf{m}$ results in additional planar Hall resistance, analogized to the local magnetization $M\mathbf{m}$. The magnitude of planar Hall resistance, $R_{\text{PHE}}$ due to the local magnetic moment $M\mathbf{m}$ is parabolic with respect to magnitude of the local magnetization $M$ via a coefficient $k$, $R_{\text{PHE}} = kM^2$ [40-42]. The Hall resistance induced by $s\mathbf{m}$ as an extra magnetization should present the same behavior of that induced by $M\mathbf{m}$. The planar Hall resistance due to $M\mathbf{m}$ is expressed as $R_{\text{P}} = R_{\text{PHE}} \sin 2\varphi$, where $\varphi$ is the azimuthal angle of magnetization $M\mathbf{m}$ [7, 12, 43, 44]. Analogically, the planar Hall resistance, $r_{\text{p}}$ due to the extra magnetization, $s\mathbf{m}$, should follow a similar trend as $r_{\text{P}} = r_{\text{PHE}} \sin 2\varphi$ and $r_{\text{PHE}} = ks^2$. Obtaining $k$ from the expression of $R_{\text{PHE}}$, we derive the expression of $r_{\text{PHE}}$ as $r_{\text{PHE}} = \dfrac{R_{\text{PHE}}}{M^2} s^2$. As such, $r_{\text{PHE}}$ can be used to calculate the magnitude of spin accumulation.

Applying a biasing DC increases the magnitude of $r_{\text{PHE}}$ to measureable levels. When AC and DC are applied in the wire concurrently, the harmonic Hall voltage induced by $s\mathbf{m}$ can be written as $v_{s,\text{Hall}} = \left[\left(\dfrac{R_{\text{PHE}}}{M^2} s^2\right) \sin 2\varphi\right]\left(j_{\text{AC}} \sin \omega t + j_{\text{DC}}\right)$, where $j_{\text{AC}}$ and $\omega$ are the amplitude and frequency of AC density respectively, and $j_{DC}$ is the magnitude of DC density. At steady-state which is the rate of spin decay equaling to that of spin generating, $s$ is proportional to the current density and can be written as $s = \zeta(j_{AC}\sin\omega t + j_{DC})$, where $\zeta$ is the coefficient constant. Substituting $s$ by $\zeta(j_{AC}\sin\omega t + j_{DC})$ in $v_{s,Hall}$ gives

$$v_{s,Hall} = R_{PHE}\frac{\zeta^2}{M^2}\sin 2\varphi\left[\left(j_{DC}^3 + 3j_{DC}^2 j_{AC}\sin\omega t + 3j_{AC}^2 j_{DC}\sin^2\omega t + j_{AC}^3\sin^3\omega t\right)\right]. \quad (1)$$

In Eq. (1), we substitute $\sin^2\omega t$ with $\frac{1}{2} - \frac{\cos 2\omega t}{2}$, as such, eliminate the constant $\frac{1}{2}$ item to obtain a second harmonic Hall voltage as $v_{s,2ndHall} = \frac{3}{2}R_{PHE}\sin 2\varphi\frac{\zeta^2}{M^2}j_{AC}^2 j_{DC}\cos 2\omega t$.

Consequently, $s\mathbf{m}$ induces a second harmonic Hall resistance $r_\alpha^{\pm\beta}$ as

$$r_\alpha^{\pm\beta} = \frac{v_{s,2ndHall}}{j_{AC}} = z_\alpha^{\pm\beta}\sin 2\varphi, \quad (2)$$

Where $z_\alpha^{\pm\beta} = \frac{3}{2}R_{PHE}\frac{\zeta^2}{M^2}j_{DC}j_{AC}$, $\alpha$ and $\beta$ correspond to the factors $\alpha\times 10^{10}$ Am$^{-2}$ for $j_{AC}$ and $\beta\times 10^{10}$ Am$^{-2}$ for $j_{DC}$, $\pm$ indicates the sign of DC. Compared with the expression of $r_{PHE} = \frac{R_{PHE}}{M^2}s^2$, the expression $z_\alpha^{\pm\beta} = \frac{3}{2}R_{PHE}\frac{\zeta^2}{M^2}j_{DC}j_{AC}$ is the other expression of $r_{PHE}$ which includes the electric current. Similarly, Eq. (2) is the other expression of $r_P$. For a constant amplitude of $j_{AC}$, the resistance $z_\alpha^{\pm\beta}$ is proportional to the amplitude of applied DC. As such, $j_{DC}$ provides a way to modulate the resistance from the baseline provided by $j_{AC}$.

This resistance $r_\alpha^{\pm\beta}$ can be obtained, by subtracting the second harmonic Hall resistance $\mathfrak{R}_\alpha^{\pm\beta}$ which is measured in experiment by DC-biased AC from that of $\mathfrak{R}_\alpha^0$ which is solely due to AC only. The measured second harmonic Hall resistance is induced by both $s\mathbf{m}$ and $M\mathbf{m}$ concurrently. As such, $\mathfrak{R}_\alpha^{\pm\beta}$ is the sum of $r_\alpha^{\pm\beta}$ due to $s\mathbf{m}$ and $R_\alpha^{\pm\beta}$ due to $M\mathbf{m}$, The second harmonic Hall resistance $R_\alpha^{\pm\beta}$ due to $M\mathbf{m}$ is

$$R_\alpha^{\pm\beta} = R_{\text{AHE}} \frac{H_{D,AC}}{2H_\perp} \cos\varphi + R_{\text{PHE}} \frac{H_{F,AC}}{H_{x-\text{ext}}} \left[2\cos^4\varphi - \cos^2\varphi\right], \quad (3)$$

where $H_\perp$ is the effective perpendicular anisotropy field, $R_{\text{AHE}}$ is the amplitude of anomalous Hall effect (AHE) resistance [7, 12, 43, 44, 45]. DC has no effect on $R_\alpha^{\pm\beta}$, alternatively, $R_\alpha^{\pm\beta} = R_\alpha^0$ [Appendix]. Both $\mathbf{H}_{D,AC}$ and $\mathbf{H}_{F,AC}$ are only determined by the AC component of electric current, $\mathbf{H}_{D,AC}$ is along $z$-axis while $\mathbf{H}_{F,AC}$ is along $y$-axis, for the wires with IMA [45]. Hence, according to Eq. (2), the measured second harmonic Hall resistance by AC only, is $\mathfrak{R}_\alpha^0 = R_\alpha^0 + 0$. While, the measured second harmonic Hall resistance by DC-biased AC is $\mathfrak{R}_\alpha^{\pm\beta} = R_\alpha^0 + r_\alpha^{\pm\beta}$, where $\beta$ is not equal to 0. Therefore, subtraction of $\mathfrak{R}_\alpha^0$ from $\mathfrak{R}_\alpha^{\pm\beta}$, $\Delta\mathfrak{R}_\alpha^{\pm\beta}(\mathfrak{R}_\alpha^{\pm\beta} - \mathfrak{R}_\alpha^0)$, equals to $r_\alpha^{\pm\beta}$. Based on Eq. (2), we conclude $\Delta\mathfrak{R}_\alpha^{\pm\beta} = \frac{3}{2} R_{\text{PHE}} \frac{\zeta^2}{M^2} j_{\text{DC}} j_{\text{AC}} \sin 2\varphi$. As such, theoretically, the coefficient indicating the magnitude of spin accumulation can be quantified by $\Delta\mathfrak{R}_\alpha^{\pm\beta}$.

Measurements of the second harmonic Hall resistances $\mathfrak{R}_\alpha^{\pm\beta}$ with respect to the azimuthal angle of magnetization were carried out in magnetic wire with stacks of Ta(8 nm)/Co(2 nm)/Pt(5 nm). The wire has IMA as evidenced by hysteresis loop measurements using both Kerr and anomalous Hall effects [45]. For all measurements, a lock-in amplifier was used to obtain the harmonic Hall voltage signals. The second harmonic Hall resistance $\mathfrak{R}_\alpha^{\pm\beta}$ is calculated by dividing the second harmonic Hall voltage with the magnitude of the AC. Only the Hall resistance modulation has been considered removing the offset resistance for each measurement. Each measured $\mathfrak{R}_\alpha^{\pm\beta}$ as well as the following $\Delta\mathfrak{R}_\alpha^{\pm\beta}$ has been moved to be around 0 Ω by eliminating a constant offset for easy comparison. A schematic of the measurement setup is shown in Fig. 1(c). The azimuthal angle of magnetization of the wire depends on the applied fields as $\varphi = \arctan\frac{H_{y-\text{ext}}}{H_{x-\text{ext}}}$, where transvers $H_{y\text{-}ext}$ sweeps from −1800 Oe to +1800 Oe along y-axis while $\pm H_{x\text{-}ext}$ keeps ±560 Oe to orientate $M\mathbf{m}$ along ±x-axis. In the following, we do not distinguish $H_{y\text{-}ext}$ from $\varphi$ as $\varphi$ is equivalent to $H_{y\text{-}ext}$, since the $\varphi$ corresponds to unique $H_{y\text{-}ext}$ for constant $H_{x\text{-}ext}$. The longitudinal magnetic field $H_{x\text{-}ext}$ was used to ensure a uniform magnetization along the wire axis [45].

The second harmonic Hall resistances, $\mathfrak{R}_4^{0,\pm 6}$, with respect to the azimuthal angle of magnetization, measured at $H_{x\text{-}ext} = -560$ Oe are presented in Fig. 1(d). For AC-$J\mathbf{s}$ only, the derived equation to represent the second Hall resistance is given in Eq. 3. Through substituting $H_\perp$, $R_{AHE}$ and $R_{PHE}$ in Eq. 3 with experimental values of 5790 Oe, 26 mΩ and 6 mΩ, respectively, the measured $R_4^0$ is in good agreement with Eq. 3. A dampinglike term $H_{D,AC}$ of 29 Oe and fieldlike term $H_{F,AC}$ of 4 Oe are obtained, as shown in Fig. 1(d). This good agreement suggests that AC-$J\mathbf{s}$ $\left(H_{F,AC}\mathbf{p} + H_{D,AC}\mathbf{m}\times\mathbf{p}\right)$ results in the symmetric behavior of $\mathfrak{R}_4^0$ with respect to $H_{y\text{-}ext}$. For the second harmonic Hall resistance obtained with using both DC and AC concurrently, an asymmetric behavior around $H_{y\text{-}ext} = 0$ Oe is observed for $\mathfrak{R}_4^{\pm 6}$. $\mathfrak{R}_4^{+6}$ and $\mathfrak{R}_4^{-6}$ are mirror symmetric to each other at $H_{y\text{-}ext} = 0$ Oe. These variations indicate that both magnitude and sign of the DC bias in the wire contributes to the corresponding signals, $\mathfrak{R}_4^{\pm 6}$. The derived equation for the second harmonic Hall resistance with DC bias is still Eq. 3 [45]. As such, we may expect to include DC induced $H_{D,DC}$ and $H_{F,DC}$ in Eq. 3 as offsets of $H_\perp$ and $H_{y\text{-}ext}$ to explain the behavior of $\mathfrak{R}_4^{\pm 6}$. However, the revised Eq. 3 fails to fit $\mathfrak{R}_4^{\pm 6}$ with fitting root-mean-square error (RMSE) reaches minimum as shown in Fig. 1(d), as the measured $\mathfrak{R}_4^{\pm 6}$ and the fitted $\mathfrak{R}_4^{\pm 6}$ by Eq. 3 do not overlap. The failure clarifies the negligible role of DC-$J\mathbf{s}$ in the behavior of $\mathfrak{R}_4^{\pm 6}$.

The differences, $\Delta\mathfrak{R}_4^{\pm 6}$, are explored to investigate the behavior of $\mathfrak{R}_4^{\pm 6}$ with respect to $H_{y\text{-}ext}$. $\Delta\mathfrak{R}_\alpha^{\pm\beta}$ is computed by subtracting the second harmonic Hall resistance obtained without DC bias ($R_\alpha^0$) from that with DC bias as $\Delta\mathfrak{R}_\alpha^{\pm\beta} = \mathfrak{R}_\alpha^{\pm\beta} - \mathfrak{R}_\alpha^0$, consequently, $\Delta\mathfrak{R}_4^{\pm 6} = \mathfrak{R}_4^{\pm 6} - \mathfrak{R}_4^0$. As shown in Fig. 1(d), $\Delta\mathfrak{R}_4^{\pm 6}$ adapts the behavior similar to the first harmonic Hall resistance with respect to $H_{y\text{-}ext}$ as shown in Fig. 1(f). The analytical expression of the first harmonic Hall resistance, which is mainly from PHE, is $R_{1stHall} = R_{PHE} \sin 2\varphi$. We use $\Delta\mathfrak{R}_4^{\pm 6} = Z_4^{\pm 6} \sin 2\varphi$ to fit $\Delta\mathfrak{R}_4^{\pm 6}$, where $2Z_4^{\pm 6}$ equals to the difference between maximum and minimum values of $\Delta\mathfrak{R}_4^{\pm 6}$, 60 μΩ. The experimental $\Delta\mathfrak{R}_4^{\pm 6}$ are in good agreement with $Z_4^{\pm 6} \sin 2\varphi$ as shown in Fig. 1(d). $R_{1stHall}$ is due to the magnetization of the wire. Analogically, $\Delta\mathfrak{R}_\alpha^{\pm\beta}$ is due to an extra magnetization and

$$\Delta\mathfrak{R}_\alpha^{\pm\beta} = Z_\alpha^{\pm\beta} \sin 2\varphi, \tag{4}$$

where $Z_\alpha^{\pm\beta}$ is the amplitude of $\Delta\mathfrak{R}_\alpha^{\pm\beta}$. $\mathfrak{R}_\alpha^0$ is given by Eq. 3. Respective of the expression of $\Delta\mathfrak{R}_\alpha^{\pm\beta} = \mathfrak{R}_\alpha^{\pm\beta} - \mathfrak{R}_\alpha^0$, $\mathfrak{R}_\alpha^{\pm\beta}$ is determined by both the pre-known AC-$J\mathbf{s}$ and the extra magnetization.

In the following, we confirm the extra magnetization is $s\mathbf{m}$, as we further show that $\Delta\mathfrak{R}_\alpha^{\pm\beta}$ follows $r_\alpha^{\pm\beta}$ with respect to the orientation of $M\mathbf{m}$, $Z_\alpha^{\pm\beta}$ is equal to the derived $z_\alpha^{\pm\beta}$ of Eq. (2), $Z_\alpha^{\pm\beta}$ with respect to $J_{DC}$ and $J_{AC}$ follows the predicted behavior of or

$$z_\alpha^{\pm\beta} = \frac{3}{2}R_{PHE}\frac{\zeta^2}{M^2}j_{DC}j_{AC}$$ in experiments.

The proposal is validated through analyzing $\Delta\mathfrak{R}_\alpha^{\pm\beta}$ obtained with varying magnetic vector of $\mathbf{H}_{x\text{-}ext}$. Figure 2(a) shows $\Delta\mathfrak{R}_4^{\pm 6}$ measured with $H_{x\text{-}ext}$ = +560 Oe. As $H_{y\text{-}ext}$ varies from −1800 Oe to +1800, the magnitude of $\Delta\mathfrak{R}_4^{+6}$ changes from −20 µΩ to +20 µΩ. In Fig. 1(d) which represents $\Delta\mathfrak{R}_4^{+6}$ with $H_{x\text{-}ext}$ = −560 Oe, $\Delta\mathfrak{R}_4^{+6}$ changes from +20 µΩ to −20 µΩ. The change of sign for $\Delta\mathfrak{R}_\alpha^{\pm\beta}$ is presented in the $\Delta\mathfrak{R}_4^{+6}$, as well as in $\Delta\mathfrak{R}_4^{-6}$. $\Delta\mathfrak{R}_\alpha^{\pm\beta}$ follow $R_{1stHall}$ or $r_\alpha^{\pm\beta}$ to change their signs when $\mathbf{H}_{x\text{-}ext}$ is orientated along opposite directions. An alternative approach to substantiate the proposal is that $\Delta\mathfrak{R}_\alpha^{\pm\beta}$ should follow $R_{1stHall}$ or $r_4^{\pm 6}$ to present extremum at $H_{y\text{-}ext} = \pm H_{x\text{-}ext}$. In experiments, the extremums of $R_{1stHall}$, ±6 mΩ, are at $H_{y\text{-}ext}$ = ±360 Oe and ±1000 Oe, when the applied constant field $H_{x\text{-}ext}$ equals to +360 Oe and +1000 Oe, respectively, as shown in Fig. 2(b). Similarly, the extremum values of $\Delta\mathfrak{R}_4^{+6}$, which are ±30 µΩ, are at $H_{y\text{-}ext}$ = ±360 Oe as shown in Fig. 2(c) where $H_{x\text{-}ext}$ is applied as +360 Oe and at ±1000 Oe as shown in Fig. 2(d) where $H_{x\text{-}ext}$ is applied as +1000 Oe.

As such, experimentally on condition that $Z_\alpha^{\pm\beta}$ follows a linear function of $j_{DC}$ and $j_{AC}$ as predicted by $\frac{3}{2}R_{PHE}\frac{\zeta^2}{M^2}j_{DC}j_{AC}$, it is allowed to conclude the extra magnetization is from the spin accumulation $s\mathbf{m}$. For different DC biases, $\mathfrak{R}_4^{\pm[1\text{ to }5]}$ were measured to compute $\Delta\mathfrak{R}_4^{\pm[1\ to\ 5]}$. $\mathfrak{R}_4^{\pm 3}$ and $\Delta\mathfrak{R}_4^{\pm 3}$ are exhibited in Fig. 3(a) with $H_{x\text{-}ext} = -560$ Oe. $\mathfrak{R}_4^{\pm 3}$ and $\Delta\mathfrak{R}_4^{\pm 3}$ show the behavior similar to $\mathfrak{R}_4^{\pm 6}$ and $\Delta\mathfrak{R}_4^{\pm 6}$, respectively. $2Z_4^{\pm 3}$ is calculated to be 30 μΩ less than $2Z_4^{\pm 6}$. Figure 3(b) shows all the computed $2Z_4^{\pm[1\ to\ 5]}$. We find that $2Z_4$ is as a linear function of the DC density. For different AC biases, $\mathfrak{R}_{6,5,4,3,2,1}^{\pm 4}$ were measured to compute $\Delta\mathfrak{R}_{6,5,4,3,2,1}^{\pm 4}$. $\mathfrak{R}_4^{\pm 4}$ and $\Delta\mathfrak{R}_4^{\pm 4}$ as examples are presented, as shown in Fig. 3(c) with $H_{x\text{-}ext} = -560$ Oe. $\mathfrak{R}_4^{\pm 4}$ and $\Delta\mathfrak{R}_4^{\pm 4}$ show the behavior similar to $\mathfrak{R}_4^{\pm 6}$ and $\Delta\mathfrak{R}_4^{\pm 6}$, respectively. Figure 3(d) shows the computed resistances $2Z_{6,5,4,3,2,1}^{\pm 4}$. $2Z_4^{\pm 4}$ is calculated to be 44 μΩ lower than $2Z_6^{\pm 4}$ of 62 μΩ. $2Z^{\pm 4}$ is as a linear function of the AC density. Hence, we conclude the extra magnetization is $s\mathbf{m}$, and $\Delta\mathfrak{R}_\alpha^{\pm\beta} = r_\alpha^{\pm\beta}$ as well as

$$Z_\alpha^{\pm\beta} = z_\alpha^{\pm\beta} = \frac{3}{2} R_{\text{PHE}} \frac{\zeta^2}{M^2} j_{\text{DC}} j_{\text{AC}}. \tag{5}$$

The coefficient, $\zeta$, which indicates the capability of electric current inducing spin accumulation can be extracted from the measurement. $\zeta^2$ is proportional to $Z_\alpha^{\pm\beta}$ as shown in Eq. (5). $Z_\alpha^{\pm\beta}$ is extracted from Hall resistances measured at various combinations of AC and DC current densities. For each AC current density, the extracted $Z$ shows a linear behavior with respect to the DC current density. Through carrying out partial derivative of $Z_\alpha^{\pm\beta}$ over $j_{\text{DC}}$ for each AC density, $\dfrac{\partial Z_\alpha^{\pm\beta}}{\partial j_{\text{DC}}}$ is obtained as shown in the inset of Fig. 4(a). $\dfrac{\partial Z_\alpha^{\pm\beta}}{\partial j_{\text{DC}}}$ is a linear function to $j_{\text{AC}}$. The slope of $\dfrac{\partial Z_\alpha^{\pm\beta}}{\partial j_{\text{DC}}}$ with respect to or as a function of $j_{\text{AC}}$, $\dfrac{\partial}{\partial j_{\text{AC}}}\left(\dfrac{\partial Z_\alpha^{\pm\beta}}{\partial j_{\text{DC}}}\right)$ is calculated to be 1.3 μΩ·[$10^{10}$ Am$^{-2}$]$^{-2}$. We obtain $\frac{\partial}{\partial j_{AC}}\left(\frac{\partial Z_\alpha^{\pm\beta}}{\partial j_{DC}}\right) = \frac{3}{2} R_{PHE} \frac{\zeta^2}{M^2}$ from Eq. (5). By substituting $R_{PHE}$ and $M$ with 5.7 m$\Omega$ and 458 emu·cc [45], respectively, $\zeta$ is computed to be 56 emu/cc per $10^{11}$ Am$^{-2}$. For the sample stack with $t=4$, $\frac{\partial Z_\alpha^{\pm\beta}}{\partial j_{DC}}$ is shown in Fig. 4(b). $\frac{\partial}{\partial j_{AC}}\left(\frac{\partial Z_\alpha^{\pm\beta}}{\partial j_{DC}}\right)$ is obtained as 0.8 $\mu\Omega\cdot[10^{10}$ Am$^{-2}]^{-2}$, similarly, $\zeta$ is computed to be 45 emu/cc per $10^{11}$ Am$^{-2}$, with $R_{PHE}$=5.6 m$\Omega$ and $M$=466 emu/cc [45]. For the sample stacks with $t=2, 6, 10$, $\frac{\partial Z_\alpha^{\pm\beta}}{\partial j_{DC}}$ is obtained at $j_{AC}=4\times10^{10}$ Am$^{-2}$ as shown in Fig. 4(c). The expression of $\frac{\partial Z_\alpha^{\pm\beta}}{\partial j_{DC}}$ is $\frac{\partial Z_\alpha^{\pm\beta}}{\partial j_{DC}} = \frac{3}{2} R_{PHE} \frac{\zeta^2}{M^2} j_{AC}$. Therefore, substituting $j_{AC}$, $R_{PHE}$ and $M$ with the experimental values in Fig. 4(c), we obtain $\zeta$ for the sample stacks with $t=2, 6, 10$, as shown in Fig. 4(d) where $\zeta$ of $t=4$ and 8 are also included. For samples with $t\leq6$, $\zeta$ is ~47 emu/cc per $10^{11}$ Am$^{-2}$. For samples with $t>6$, $\zeta$ increases from 56 emu/cc per $10^{11}$ Am$^{-2}$, reaching a maximum of 107 emu/cc per $10^{11}$ Am$^{-2}$ at $t = 10$.

The critical current density for SOT induced magnetization switching and domain wall driving is in the order of $1\times10^{11}$ Am$^{-2}$. The total current density in our experiment is in the same order for wires investigated. For films with $t=2, 4, 6, 8$, and 10, the measured spin accumulation is 54 emu/cc, 45 emu/cc, 44 emu/cc, 56 emu/cc and 107 emu/cc, respectively. The corresponding local magnetization is 581 emu/cc, 446 emu/cc, 436 emu/cc, 458 emu/cc, and 592 emu/cc. Therefore, the ratio of the spin accumulation $s$ to the local magnetization $M$, is ~10% for the films with $t=2, 4$ and 6, 12% for $t=8$, and 18% for 10. The percentage indicates that the spin density induced by a current density of $1\times10^{11}$ Am$^{-2}$ is in the comparable order to the local magnetization in all the stacks, as we firstly claim. As the initial orientation of spin accumulation is along *y*-axis, when the local magnetization orientates along *y*-axis, the spin accumulation can vary the magnetization by as much as 13% in the wire with $t$=10. Therefore, to switch magnetization and drive domain wall motion by a current density of $1\times10^{11}$ Am$^{-2}$ is within expectation. If current density is increased to $1\times10^{12}$ Am$^{-2}$ which is the critical current for spin-transfer torque to switch magnetization and drive domain wall motion, the spin accumulation can be 540 emu/cc, 450 emu/cc, 440 emu/cc, 560 emu/cc and 1070 emu/cc for films with $t$=2, 4, 6, 8, and 10, respectively. The magnitudes of spin accumulation are close to the magnitudes of the corresponding local magnetization. In this case, the magnetic moments can be reorganized in the wires. Hence, the orientation of magnetization could be determined by the spin accumulation.

As shown in Fig. 4(d), $\zeta$ follows the spin Hall angles of the samples which have been reported in our previous work with respect to the thickness of Ta layer [45], irrespective of their magnitudes. $\zeta$ represents the extra magnetization generated by an electric current, while spin Hall angle of Ta/Pt indicates the percentage of spin current converted by Ta/Pt in an electric current. As such, the same trend of $\zeta$ and spin Hall angle with respect to the thickness of Ta is expected, since the extra magnetization should be as a linear function to magnitude of spin current. Therefore, the same trend confirms the spin accumulation in Co layer is from the Ta and Pt layers.

**Conclusion**

We have experimentally quantified the spin accumulation induced by electric current in series stacks of Ta/Co/Pt. We find that spin accumulation is around dozens of percentage of local magnetization when the current density is $10^{11}$ Am$^{-2}$. As our results demonstrate for the first time, when the spins from Ta and Pt are in relaxation state, they still contribute to second harmonic Hall resistance, instead when they are in initial state only as expected. The coefficient of spin accumulation over electric current is consistent with spin Hall angle. This consistency suggests that the coefficient can be used to evaluate the capability of a heavy metal converting electric current to spin current. As the measurements are easily carried out, we offer a concise solution to estimate spin accumulation.

**Acknowledgements**

This work was supported by the Singapore National Research Foundation, Prime Minister's Office, under a Competitive Research Programme (Non-volatile Magnetic Logic and Memory Integrated Circuit Devices, NRF-CRP9-2011-01), and an Industry-IHL Partnership Program (NRF2015-IIP001-001). The work was also supported by a MOE-AcRF Tier 2 Grant (MOE 2013-T2-2-017). WSL is a member of the Singapore Spintronics Consortium (SG-SPIN).

**Figures and captions**

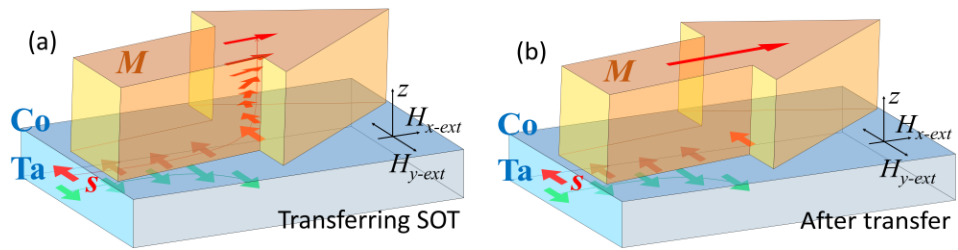

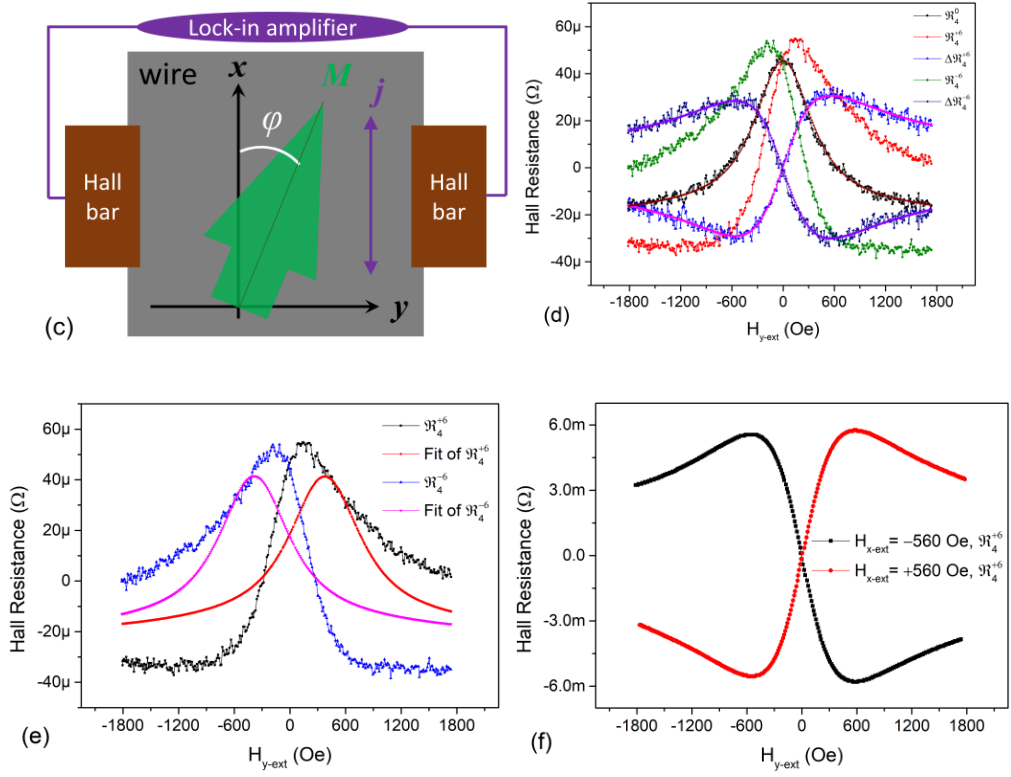

Fig. 1 (a) the transfer of spin torque from *s* to *M* in dozens of nanoseconds; (b) *s* relaxes to be along M after SOT transferring; (c) schematic of measurement setup; (d) measured $\mathfrak{R}_4^{\pm 6}$ and obtained $\Delta\mathfrak{R}_4^{\pm 6}$ with respect to $H_{y\text{-}ext}$, fit of $\Delta\mathfrak{R}_4^{\pm 6}$ indicate that $\Delta\mathfrak{R}_4^{\pm 6}$ is fitted by Eq. (2) where $H_{x\text{-}ext}$ is 560 Oe; (e) measured $\mathfrak{R}_4^{\pm 6}$ with respect to $H_{y\text{-}ext}$, fit of $\Delta\mathfrak{R}_4^{+6}$ indicate that $\Delta\mathfrak{R}_4^{+6}$ is fitted to reaching minimum RMSE by Eq. (1) with changing $H_{x\text{-}ext}$ to 560+69 Oe and $H_{y\text{-}ext}$ to 560+38 Oe, fit of $\Delta\mathfrak{R}_4^{-6}$ indicate that $\Delta\mathfrak{R}_4^{-6}$ is fitted to reaching minimum RMSE by Eq. (1) with changing $H_{x\text{-}ext}$ to 560+60 Oe and $H_{y\text{-}ext}$ to 560−39 Oe; (f) measured first harmonic Hall resistances with respect to $H_{y\text{-}ext}$ when obtaining $\mathfrak{R}_4^{+6}$ with $H_{x\text{-}ext} = \pm 560$ Oe. In (d), the magenta and violet lines show the fit to the experimental data.

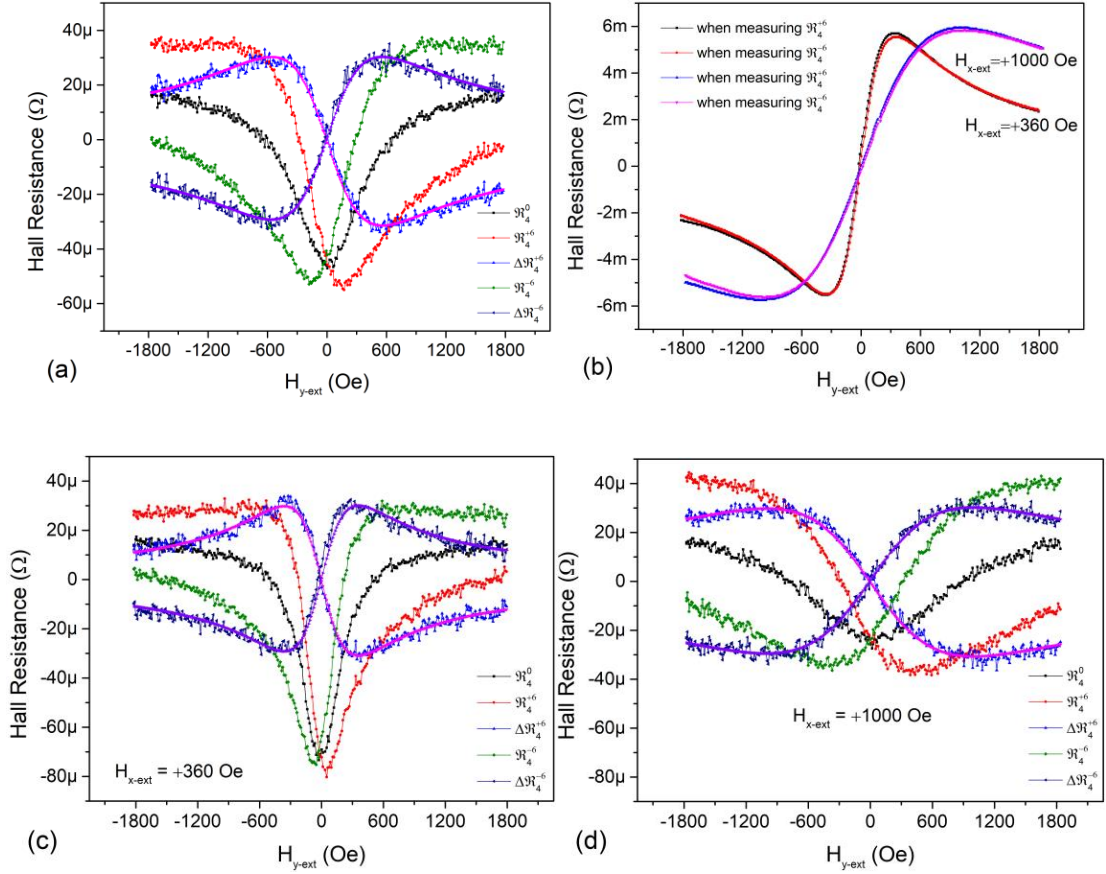

Fig. 2 (a) measured $\mathfrak{R}_4^{\pm 6}$ and obtained $\Delta\mathfrak{R}_4^{\pm 6}$ with respect to $H_{y\text{-}ext}$ when $H_{x\text{-}ext}=-560$ Oe; (b) obtained first harmonic Hall resistance with respect to $H_{y\text{-}ext}$ when measuring corresponding $\mathfrak{R}_4^{\pm 6}$ with $H_{x\text{-}ext}$ equating to +360 Oe and +1000 Oe; (c) measured $\mathfrak{R}_4^{\pm 6}$ and obtained $\Delta\mathfrak{R}_4^{\pm 6}$ with respect to $H_{y\text{-}ext}$ when $H_{x\text{-}ext}=+360$ Oe; (d) measured $\mathfrak{R}_4^{\pm 6}$ and obtained $\Delta\mathfrak{R}_4^{\pm 6}$ with respect to $H_{y\text{-}ext}$ when $H_{x\text{-}ext}=+1000$ Oe. In (a), (c) and (d), the magenta and violet lines show the fit to the experimental data.

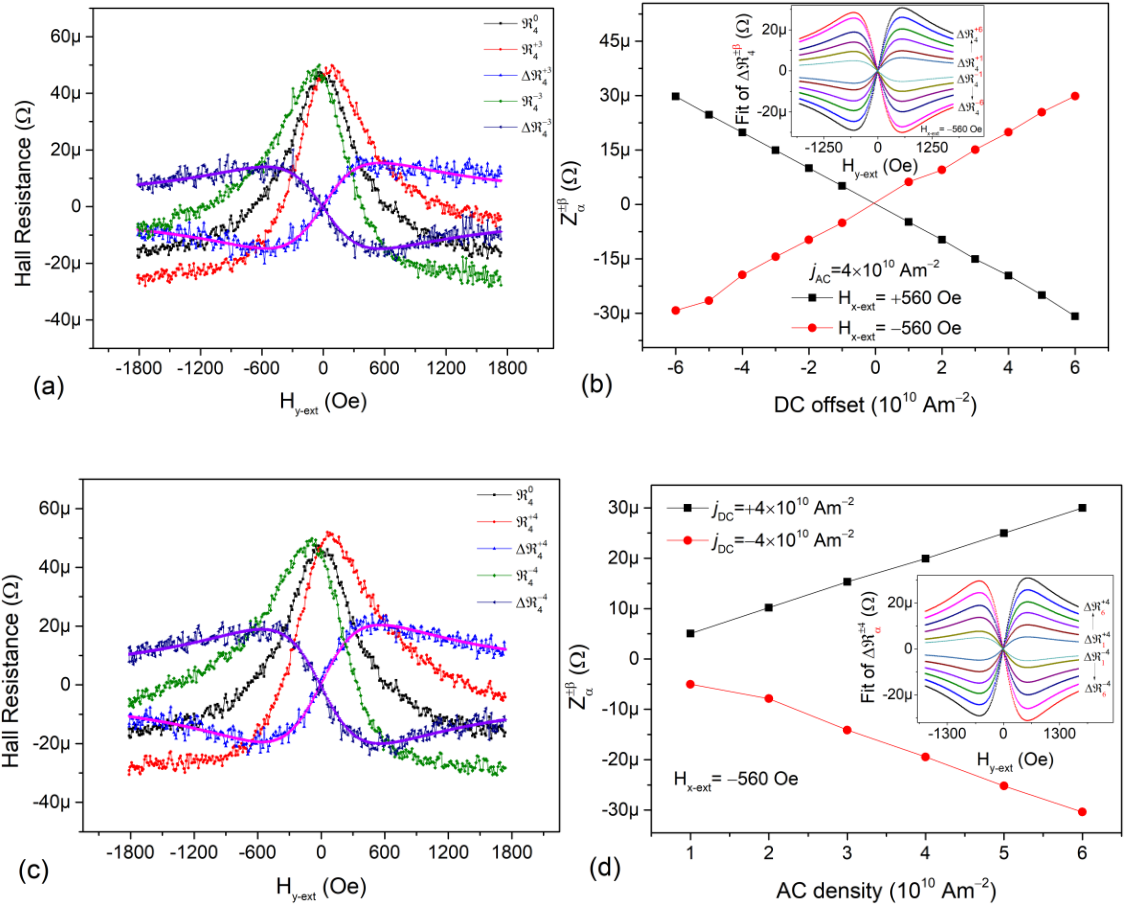

Fig. 3 (a) measured $\mathfrak{R}_4^{\pm 3}$ and obtained $\Delta\mathfrak{R}_4^{\pm 3}$ with respect to $H_{y\text{-}ext}$ when $H_{x\text{-}ext}=-560$ Oe; (b) calculated $Z_4^{\pm\beta}$ with respect to the DC offset $\beta$ when the AC density is fixed to be $4\times 10^{10}$ Am$^{-2}$; (c) measured $\mathfrak{R}_4^{\pm 4}$ and obtained $\Delta\mathfrak{R}_4^{\pm 4}$ with respect to $H_{y\text{-}ext}$ when $H_{x\text{-}ext}=-560$ Oe; (d) calculated $Z_\alpha^{\pm 4}$ with respect to the AC $\alpha$ when the DC densities are fixed to be $\pm 4\times 10^{10}$ Am$^{-2}$. In (a) and (c), the magenta and violet lines show the fit to the experimental data.

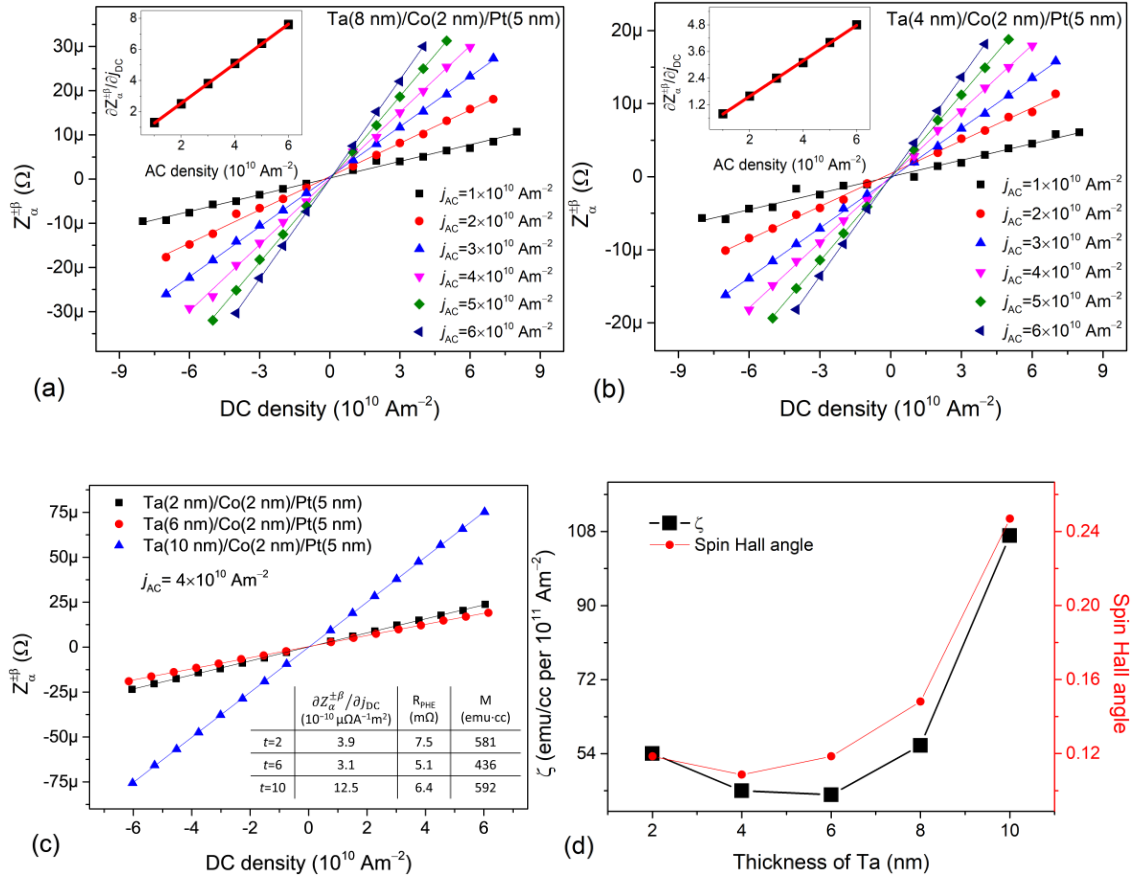

Fig. 4 calculated $Z_\alpha^{\pm\beta}$ with respect to the DC offset $\beta$ under different AC densities, inset is the slope of $Z_\alpha^{\pm\beta}$ to $\beta$ with respect to $\alpha$ for sample (a) Ta(8 nm)/Co(2 nm)/Pt(5 nm) and sample (b) Ta(4 nm)/Co(2 nm)/Pt(5 nm); (c) calculated $Z_4^{\pm\beta}$ with respect to the DC offset $\beta$ for the samples of Ta(*t* nm)/Co(2 nm)/Pt(5 nm) with *t*=2, 6 and 10; (d) calculated $\zeta$ (black line) and reported spin Hall angle (red line) with respect to the thickness of Ta.